\documentclass[aps,prd,twocolumn,floatfix]{revtex4}
\usepackage{graphicx}
\usepackage{amsmath}
\usepackage{amssymb}
\usepackage{epsfig}
\usepackage{color}
\usepackage{textcomp}
\usepackage{color}

\newcommand{\be}{\begin{equation}}
\newcommand{\ee}{\end{equation}}

\newcommand{\revis}{}

\begin{document}

\title{Chirped-Frequency Excitation of Gravitationally Bound Ultracold Neutrons}
\author{Giovanni Manfredi}
\email{giovanni.manfredi@ipcms.unistra.fr}
\affiliation{Institut de Physique et Chimie des Mat\'{e}riaux de
Strasbourg, CNRS and Universit\'{e} de Strasbourg, BP 43, F-67034 Strasbourg, France}
\author{Omar Morandi}
\affiliation{Institut de Physique et Chimie des Mat\'{e}riaux de
Strasbourg, CNRS and Universit\'{e} de Strasbourg, BP 43, F-67034 Strasbourg, France\\ \\
Dipartimento di Matematica e Informatica "U. Dini", Università di Firenze, 50139 Firenze, Italy
}
\author{Lazar Friedland}
\affiliation{Racah Institute of Physics, Hebrew University of Jerusalem, Jerusalem 91904, Israel}
\author{Tobias Jenke}
\affiliation{Institut Laue-Langevin, 71 avenue des Martyrs, 38000 Grenoble, France\\ \\
Atominstitut, Technische Universit\"at Wien, Stadionallee
2, 1020 Vienna, Austria}
\author{Hartmut Abele}
\affiliation{Atominstitut, Technische Universit\"at Wien, Stadionallee
2, 1020 Vienna, Austria}

\begin{abstract}
Ultracold neutrons confined in the Earth's gravitational field display quantized energy levels that have been observed for over a decade. In recent resonance spectroscopy experiments [T. Jenke et al., Nature Phys. {\bf 7}, 468 (2011)], the transition between two such gravitational quantum states was driven by the mechanical oscillation of the plates that confine the neutrons.
Here we show that, by applying a sinusoidal modulation with slowly varying frequency (chirp), the neutrons can be brought to
higher excited states
by climbing the energy levels one by one.
The proposed experiment should make it possible to observe the quantum-classical
transition that occurs at high neutron energies.
Furthermore, it provides a technique to realize superpositions of
gravitational quantum states, to be used
for precision tests of gravity at short distances.

\end{abstract}

\maketitle

% PACS
% 04.80.Cc 	Experimental tests of gravitational theories
% 37.90.+j 	Other topics in mechanical control of atoms, molecules, and ions
% 05.45.-a 	Nonlinear dynamics and chaos
% 03.75.Dg 	Atom and neutron interferometry

\section{Introduction}
Gravity is a very weak force and thus very difficult to test except on astronomical scales, where the effect of other interactions becomes negligible. Even there, current theories fail to explain a number of observed phenomena, such as the rotation speed of galaxies, without postulating the existence of unknown forms of matter {\revis or types of interactions}.
Gravity's behavior on small scales and its relation to quantum mechanics are still not fully understood~\cite{quantgrav}. However, in the last decade, several experiments and theoretical developments have helped clarify how a quantum object responds to gravity, in particular the gravitational field of the Earth. These include
{\revis the measurement of gravitationally induced quantum phases by neutron interferometry \cite{COW_1975}},
interference experiments on free-falling Bose-Einstein condensates (BECs) \cite{Bouncing-BEC}, the observation
of the quantized energy levels of ultracold neutrons (UCNs) \cite{Nesvizhevsky2002},
{\revis and the realization of gravity-resonance spectroscopy \cite{Jenke_NP11}, which was recently exploited to search
for extra short-range interactions~\cite{Jenke_PRL14, cronenberg}.
Tests of the equivalence principle in the quantum regime were also proposed \cite{equiv-principle,Kajari2010}.}
Other ongoing experiments aim at measuring the gravitational acceleration of antimatter \cite{GBAR_2012, AEGIS, ALPHA}.
%using interferometric techniques (AEGIS collaboration~\cite{AEGIS}) or by direct %observation of the free fall of antihydrogen atoms (GBAR~\cite{GBAR_2012} and  %ALPHA~\cite{ALPHA}).

Here, we focus on experiments using UCNs confined in the gravitational potential of the Earth. As a consequence of standard quantum mechanics, the energy levels of the neutrons are quantized, with a typical energy for the ground state of the order of 1 peV. The classical height~$h$ corresponding to such an energy ($mgh$, where~$m$ is the neutron mass and~$g$ is the Earth's gravitational acceleration) is of the order of $10~\rm \mu m$.

{\revis
The quantum gravitational states of UCNs were observed for the first time by Nesvizhevsky et al. \cite{Nesvizhevsky2002}, based on an earlier proposal \cite{Luschikov_Frank_1978}.
More recent studies were pursued independently by three groups:
(i) The GRANIT experiment aims at realizing transitions between gravitational quantum states using inhomogeneous magnetic fields \cite{Baessler_2011}; (ii) Ichikawa et. al. \cite{Ichikawa2014} reported on measurements of the spatial distribution of these states using a dedicated high-resolution detector; (iii) Finally, in a series of experiments conducted at the PF2 source of the Institut Laue-Langevin (qBounce project),
mechanical oscillations of the neutron mirrors were used to induce transitions between several gravitational quantum states \cite{Jenke_NP11, Jenke_PRL14, cronenberg}.
}

In the latter experiments, the UCNs first go through a collimator that selects very small vertical velocities. Then, they are injected into an apparatus (about 15~cm long) made of two parallel plates -- a mirror on the bottom and a rough, scattering mirror on the top -- separated by a narrow slit~\cite{Jenke2013}. The neutrons flow between {\revis the two plates}, and their transmission is measured at the exit as a function of the slit thickness.
The transition between two quantum states is triggered by a sinusoidal oscillation of the plates.
These experiments pave the way to the purely mechanical control and manipulation of quantum gravitational states, with potential applications to fundamental physics, such as the testing of dark matter scenarios and nonstandard theories of gravity~\cite{Jenke_PRL14,cronenberg}.
Here, we show that a chirped oscillation with slowly varying frequency (classically known as {\it autoresonance}) can be used to bring the neutrons to higher excited states by climbing the energy levels one by one. The proposed experiment may have an impact on several areas of fundamental physics. In particular, it could allow the observation of the transition from quantum ladder climbing to classical autoresonance \cite{Friedland_AJP} which occurs at high energies.
Further, it provides a viable technique to realize quantum superpositions of
gravitational states. Such superpositions may be used to perform quantum interference experiments similar to those realized for free-falling BECs \cite{Bouncing-BEC}, with the interfering objects being single neutrons instead of many-body systems such as BECs.

\section{Quantum bouncer}
The system under consideration is conceptually simple: a quantum particle falls freely in the Earth's gravitational field and bounces off a perfectly reflecting plate.
The gravitational potential~$U(z)=mgz$ is assumed to be linear {\revis in the} %with the
height~$z$ and to not depend on the other coordinates, so that the problem is essentially one-dimensional. The relevant Schr\"{o}dinger equation is then:~$i\hbar \partial_t \psi = H \psi$, with~$H = p_z^2/2m + mgz$. The wave function must vanish at both~$z \to +\infty$ and the position of the oscillating plate,~$z=L(t)=L_0 \cos \phi_d(t)$, where~$L_0$ and~$\phi_d(t)$ are the phase and amplitude of the mechanical oscillations. For a fixed surface ($L_0=0$), the eigenstates are pieces of the same Airy function~${\rm Ai}(z)$ and the corresponding eigenvalues are the zeroes of~${\rm Ai}(z)$~\cite{Banacloche,Abele_NJP12}. This system is known  as the ``quantum bouncer" or ``quantum trampoline".

For an oscillating plate, it is convenient to transform to a reference frame where the plate is fixed, by defining the coordinate~$x=z-L(t)$. In this frame, the Hamiltonian becomes~$H = p_x^2/2m + mgx - mL_0 \omega_d^2 x \cos \phi_d$, where
%the new term comes from the acceleration of the reference frame and
$\omega_d(t) = {\dot \phi}_d$ is the oscillation frequency (the dot denotes differentiation with respect to time).

Throughout this work, we shall use scaled variables where space is normalized to~$a=[\hbar^2/(2m^2g)]^{1/3}=\, 5.87 \rm \mu m$ and time to~$T=ma^2/\hbar = 0.547\,\rm ms$ (the corresponding energy and frequency are~$\mathcal{E}_0=\hbar/T = 1.20\, \rm peV$ and~$ f_0=T^{-1}= 1.83 \,\rm kHz$). In these units, the problem is fully characterized by three dimensionless quantities: the scaled driving amplitude~$\epsilon=L_0/a$, the frequency~$\omega_d T$, and the chirp rate~$\dot\omega_d T^2$.

\section{Semiclassical autoresonance theory}
Autoresonant excitation is a technique originally devised for a nonlinear oscillator driven by a chirped force with slowly varying frequency, i.e.,~$\dot \omega_d \ll \omega_d^2$ (adiabatic regime).
If the driving amplitude exceeds a certain threshold, then the nonlinear frequency of the oscillator stays locked to the excitation frequency, so that the resonant match between the drive and the oscillator is never lost, and the amplitude of the oscillations grows without limit~\cite{Friedland_AJP,Friedland_scholarpedia}.
Its quantum-mechanical limit is the so-called ladder climbing of a series of discrete quantum states~\cite{Friedland_ladderclimb,Barth2011}.

We now sketch the main steps of the autoresonance theory for the classical bouncer and extend it to the quantum regime via semiclassical arguments.
In the dimensionless units defined above, the Hamiltonian reads as
\be
H(x,p,t) = \frac{p^2}{2} +  \frac{x}{2} - \epsilon\, {\omega}_d^2(t) x \cos\phi_d(t).
\label{eq:hamilton}
\ee
This Hamiltonian can be transformed to action-angle variables~$(I,\theta)$,  to yield:~$H(I,\theta,t) = H_0(I) - \epsilon{\omega}_d^2 x(I,\theta) \cos\phi_d$, where~$H_0 = \frac{3}{2}b I^{2/3}$ and ~$b=(\pi^2/12)^{1/3}$.
The unperturbed trajectory over half a period can be written as:
$x(I,\theta)= [\pi^2- \theta^2 ]/(4\Omega^2)$,
where~$\Omega(I)=H'_0(I)=b I^{-1/3}$ is the frequency and the apex stands for differentiation with respect to~$I$.
We now expand the position in a Fourier series of the angle:~$x(I,\theta)=\sum_n a_n(I) \cos(n\theta)$ and keep only the first term
$a_1(I)= {2\over \pi}\int_{0}^{\pi} x(I,\theta)\cos\theta\,d\theta = \Omega^{-2}$
(Chirikov's single resonance approximation~\cite{Sagdeev}). The single-resonance Hamiltonian reads as
\be
H= H_0(I) -  \frac{\epsilon}{2}\,\frac{\omega_d^2}{ \Omega^2} \cos\Phi,
\label{eq:hamilton-singleres}
\ee
where~$\Phi \equiv \theta-\phi_d$ is the phase difference between the bouncer and the drive.
Autoresonance occurs when this phase difference stays bounded, so that the resonance condition holds at all times.
The corresponding Hamilton's equations read as
\begin{eqnarray}
\dot I &=& -\frac{\epsilon}{2} \frac{\omega_d^2}{\Omega^2} \sin\Phi,
\label{eq:hamilton-eq-action}
\\
\dot \Phi &=&  \Omega - \omega_d + \epsilon\frac{\Omega'}{\Omega^3}\, \omega_d^2\cos\Phi.
\label{eq:hamilton-eq-angle}
\end{eqnarray}

We now expand the action variable~$I(t)=\bar{I}+\Delta I(t)$ around~$\bar I(t)$, defined as the value for which the phase-locking between the drive and the bouncer is perfect, i.e.,~$\Omega(\bar{I})=\omega_d(t)$.
If such a phase-locking could be sustained, then the action~$I$ would be controlled by simply varying~$\omega_d$.
As, in practice, the phase-locking is only approximate, the action will perform small oscillations~$\Delta I$ around~$\bar{I}$~\cite{Friedland_AJP}.
Using the above expansion for~$I(t)$ and neglecting the last term in Eq. \eqref{eq:hamilton-eq-angle}, the equation of motion for the phase mismatch becomes
\be
\ddot\Phi = \frac{\epsilon}{2} \Omega'(\bar{I})\,\frac{\omega_d^2}{\Omega^2(\bar{I})} \sin\Phi - \dot \omega_d.
\label{eq:eqmotion}
\ee

Several interesting conclusions can be drawn from Eq. \eqref{eq:eqmotion}. First, for negligible~$\dot \omega_d$ there is a stationary state around~$\Phi=0$, which means that the oscillator and the drive can indeed be locked in phase -- this is the telltale signature of autoresonance.
Second, for finite~$\dot \omega_d$, such a stable stationary solution exists provided that
\be
|\dot \omega_d| <  \frac{\epsilon}{2}\, |\Omega'(\bar{I})|\, \frac{\omega_d^2}{\Omega^2(\bar{I})}  \approx \frac{\epsilon}{6} \frac{\omega_d^4}{b^3},
\label{eq:trapping}
\ee
where we used the fact that, in autoresonance,~$\Omega \approx {\omega}_d$. Equation \eqref{eq:trapping} shows that a lower drive amplitude~$\epsilon$ requires a slower chirp rate~$\dot \omega_d$ for the autoresonant trapping to occur.
The drive frequency must decrease with time ($\dot\omega_d <0$) because the classical bouncing period increases with height.

{\revis
Equation \eqref{eq:trapping} can be used to find an optimal chirp for which the trapping condition is never lost. Inserting on the right-hand side of the Eq. \eqref{eq:trapping} a multiplicative ``safety factor"~$q<1$, we obtain the following solution for the driving frequency~$\omega_d(t)$:
\be
\frac{1}{\omega_d^3(t)} = \frac{1}{\omega_d^3(0)} + q \,\frac{\epsilon t}{2b^3}.
\label{eq:optdrive}
\ee
Solving for~$t$, the above equation can also be used to find the time at which a certain frequency is reached.}
In the experiments it may be easier to work with a constant chirp rate, although any chirp that satisfies Eq. \eqref{eq:trapping} for a certain lapse of time can be used. Importantly, the various transitions will be excited automatically and without the need of any external control.

Finally, Eq. \eqref{eq:eqmotion} (in the adiabatic limit ~$\dot \omega_d \to 0$) is identical to the equation of a classical pendulum, with~$\Delta I=-\dot\Phi/\Omega'(\bar{I})$ playing the role of the momentum. Thus, we can estimate the width of the resonance~$\Delta I_{max}$, defined as the maximum excursion of~$\Delta I$:
\be
\Delta I_{max} = \sqrt{8\frac {\epsilon\,\omega_d^2 / \Omega^2(\bar{I})}{|\Omega'(\bar{I})|}}
= \frac{12}{\pi} \sqrt{2\epsilon}\, \omega_d\, \bar{I}
= 5\sqrt{\epsilon}\,\bar{I}^{2/3}.
\label{eq:delta_I}
\ee
where we used the fact that~$\omega_d=\Omega(\bar{I})=b\bar{I}^{-1/3}$.
Semiclassically, the action is related to the energy levels through the relation~$I_n \sim \hbar n$. Therefore, Eq. \eqref{eq:delta_I} also represents the number of quantum states involved in the dynamics, denoted~$\Delta n$, and it can be used to estimate the threshold characterizing the quantum-classical transition, which should occur when~$\Delta n$ is large.
These semiclassical considerations are in good agreement with the full quantum simulations shown below.

\section{Numerical results}
The results were obtained by solving numerically the time-dependent 1D Schr\"{o}dinger equation with the Hamiltonian \eqref{eq:hamilton}.
{\revis
In the  simulations, the UCNs are initially prepared in their ground state ($n=1$). (This is of course trickier to achieve in practice, but recent experiments managed to prepare around~$70\%$ of the neutrons in the ground state~\cite{Jenke_PRL14}). Subsequently, we start driving the system with a chirped sinusoidal modulation in order to induce the transitions.
The driving frequency is given by Eq. \eqref{eq:optdrive} with
the following dimensionless parameters: Excitation amplitude~$\epsilon=0.228$, initial frequency~${\omega}_d(0) = 1.205$, and safety factor~$q=0.5$.}
We will discuss later how these numbers relate to the physical parameters used in the experiments.

The key result is depicted in Fig. 1, where we show the occupation probabilities of the various quantum levels. Since the drive frequency decreases, time actually flows from right to left on the figure. Initially, only the ground state ($n=1$) is occupied. When the drive frequency is swept through the first resonance~$\omega_{1\to 2}=(E_2-E_1)/\hbar = 2\pi\times 254 \,\rm Hz$, the system jumps to the second state {\revis with around~$70\%$ efficiency. This first jump occurs after about 10~ms}. As the frequency continues to decrease, the system goes through all successive quantum levels one by one, although the population transfer becomes less and less efficient. After a certain time {\revis ($\omega \approx \omega_3$ in Fig. 1, corresponding to $t \approx 85~\rm ms$)} many levels are occupied simultaneously, which signals the transition to the classical regime. The occupation probabilities at three representative frequencies are plotted in the top panel of the figure, and show that the distribution gets wider and wider with time.
\begin{figure}[tb]
\begin{center}
{\includegraphics[width=.5\textwidth]{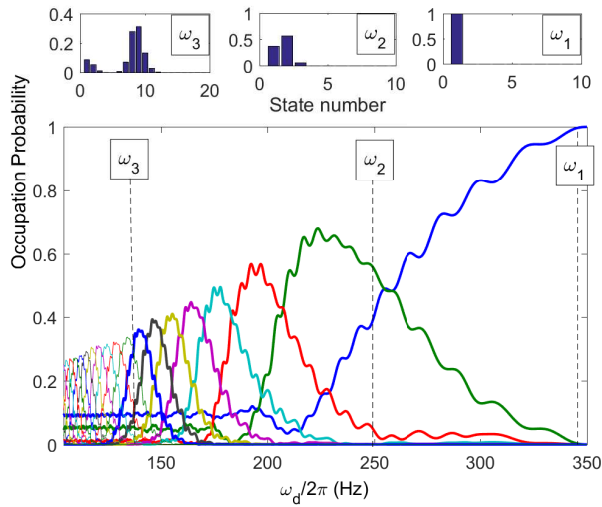}}
  \caption{{\it Color online}. Bottom panel: Occupation probabilities of the different energy levels as a function of the drive frequency~$\omega_d$. Top panel: Occupation probabilities as a function of the quantum state number at three values of the drive frequency ($\omega_1=2\pi \times 338\, \rm Hz,\,\omega_2=2\pi \times 248\, \rm Hz,\, \omega_3=2\pi \times 136\, \rm Hz$). These frequencies are shown as dashed vertical lines in the bottom panel.}
  \label{fig:occupation}
\end{center}
\end{figure}

The state of the system can be conveniently pictured using its Wigner function, which represents a pseudo-probability distribution in the classical phase space~$(x,p)$ (it is not a true probability distribution as it can take negative values).
Figure 2 shows the Wigner function corresponding to the driving frequencies~$\omega_{1,2,3}$ defined in Fig. 1. For~$\omega_d=\omega_1$ (top panel), the system is still basically in its ground state, and the Wigner distribution is a smooth function of its variables. In contrast, at~$\omega_d=\omega_3$ (bottom panel) the bouncer is already in the semiclassical regime and its Wigner function follows closely the classical trajectory~$p=\sqrt{2H-x}$ in the phase space (dashed line).
The wavy pattern observed in the Wigner function reveals that the UCNs are in a coherent superposition of many quantum states.

\begin{figure}[tb]
\begin{center}
{\includegraphics[width=.5\textwidth]{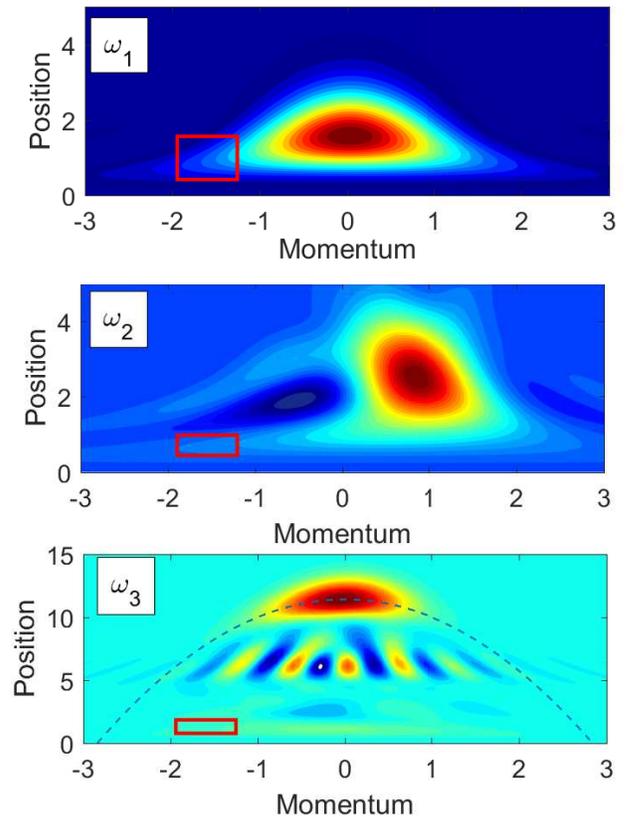}}
  \caption{{\it Color online}. Wigner functions in the phase space~$(x,p)$ at three different times corresponding to the three frequencies~$\omega_{1,2,3}$ shown in Fig. 1. The red rectangle shows a phase-space area equal to~$\hbar$. In the bottom panel, the dashed line shows the corresponding classical trajectory. Position and momentum are expressed in normalized units.}
  \label{fig:wigner}
\end{center}
\end{figure}

Finally, in Fig. 3 we show the average energy of the bouncer as a function of the driving frequency.
{\revis
After a few quantum oscillations, the system quickly approaches the classical regime, represented by the classical result~$H_0(\Omega)=\pi^2/(8\Omega^2)$ (dashed line).
}
The inset of Fig. 3 shows the average occupation number~$\langle n \rangle$, as well as the width (variance) of the level distribution~$\Delta n$, represented by the two curves~$\langle n \rangle \pm \Delta n$. As expected, the variance increases with time (i.e., with decreasing drive frequency) as more and more levels are simultaneously excited.
{\revis
For instance, for~$\omega_d =0.7 \, f_0 \approx 2\pi\times 200 \,\rm Hz$, the semiclassical estimate of Eq. \eqref{eq:delta_I} yields~$\langle n \rangle  \approx 3$ and~$\Delta n = 5$.
}
\begin{figure}[tb]
\begin{center}
{\includegraphics[width=.5\textwidth]{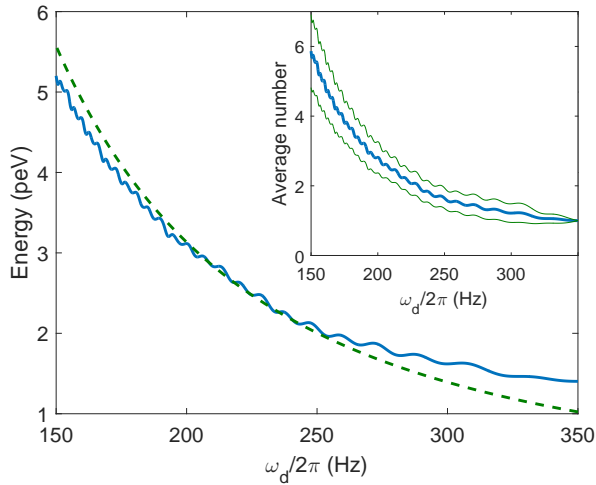}}
  \caption{{\it Color online}. Average energy of the bouncer system as a function of the drive frequency.  The green dashed line represents the classical formula. Inset: Average occupation number~$\langle n \rangle$ (thick blue line) and width of the level distribution~$\langle n \rangle \pm \Delta n$ (thin green lines) as a function of the drive frequency. }
  \label{fig:energy}
\end{center}
\end{figure}

\section{Experimental realization}
In order to compare our numerical results with those of recent experiments, we first express all parameters in physical units.
In the simulations described above, the initial driving frequency was {\revis~$\omega_d(0)=1.205 \,f_0 = 2\pi\times 350 \,\rm Hz$, }
larger than the first transition frequency~$\omega_{1\to 2}= 2\pi\times 254 \,\rm Hz$ \footnote{The computed transition frequency is very close to the frequency observed in the experiments~\cite{Jenke_PRL14}~$\omega_{1\to 2}= 2\pi\times 258.2 \,\rm Hz$.}.
During the excitation, the drive frequency was decreased adiabatically according to Eq. \eqref{eq:optdrive}.
The drive amplitude {\revis~$L_0=\epsilon a = 1.34\, \rm \mu m$} is much smaller than the size of the neutrons ground state ($\approx 13.7\, \rm \mu m$). The drive strength can also be expressed in terms of the acceleration~$\gamma=L_0 \omega_d^2$.
{\revis Initially~$\gamma(0) = 6.48\, \rm ms^{-2}$, while at the first resonance~$\gamma_{1 \to 2}= 3.41\, \rm ms^{-2}$.
These values are similar to those used in the experiments of Jenke et al.~\cite{Jenke_NP11}.}

{\revis
An important quantity is the total time~$t_{\displaystyle *}$ it takes to reach a specific energy level, for instance~$n=9$, which roughly corresponds to~$\omega_3$ on Fig. 1. Solving for~$t$ in Eq. \eqref{eq:optdrive}, we obtain~$t_{\displaystyle *} = 87 \rm ms$.
The typical horizontal velocity of the neutrons~\cite{Jenke_NP11} being of the order of~$6\rm m s^{-1}$, this would require the mirror plate to be 50~cm long, while in the most recent published
experiment it is 20~cm~\cite{cronenberg}.
This gap could be bridged with relative ease by increasing the length of the mirror and slightly reducing the transit speed of the neutrons.
Indeed, a mirror of length 34~cm is currently being developed; 80~cm-long mirrors have been proposed~\cite{abele_prd}; and a
further reduction of the neutron velocities to about~$4~\rm m s^{-1}$ could be achieved easily through a shift of the UCN energy spectrum in the Earth's gravitational potential. Another way to work with slower neutrons would be to install the experiment at a Helium-based superthermal neutron source~\cite{Zimmer_2011}, which produces a very soft velocity spectrum.
In addition, simulations using a stronger drive showed that the time to reach the level~$n=9$ could be further reduced to around 50~ms, albeit with reduced efficiency.
}

\section{Discussion}
In this work, we proposed a new technique to manipulate UCNs, which relies on the periodic modulation of a mirror plate upon which the neutrons bounce off. Past experiments, in particular those using the q\textsc{BOUNCE} setup, were performed at fixed frequency and involved the transition between two quantum states. The key technique used in these experiments is the newly-developed Gravity Resonance Spectroscopy (GRS) method \cite{Jenke_PRL14}.
In the present work, we showed that, by using a chirped modulation with slowly varying frequency, it is possible to reach much higher quantum states and potentially observe the transition from quantum to classical behavior.

This is an interesting achievement for many reasons.
One of the aspects of chirped-frequency excitations is the fact that this method provides a test of gravity at short distances \cite{Abele_2008,RMP_neutrons_cosmo,Nesvi_PRD2008} in a complementary way compared to the previous GRS experiments. Within the framework of the present proposal, ultracold neutrons are transferred into a coherent superposition of many quantum states, and not just two as in the earlier GRS experiments. After this excitation phase, it will be possible for the neutron wave function to evolve freely in time. This wave function, which is a superposition of several quantum gravitational states (see Fig. 2, bottom frame), should therefore display some level of quantum interference between the various eigenfunctions of which it is composed.
These superpositions can be measured experimentally. A spatial resolution detector for the measurement of the squared wave function -- i.e. the probability to find a neutron on the neutron mirror -- was developed in the past  by some of the present authors \cite{Jenke2013}.
Measurements of the neutron density distribution above the mirror can provide a test of Newtonian gravity at short distances. A non-Newtonian contribution would change both the wave function and the energy eigenstates, and thus the time evolution of the neutron.

Furthermore, the proposed technique will access higher quantum states, thus providing a higher sensitivity for gravity tests with respect to what is possible today. Indeed, any modification of Newtonian gravity will be felt, on the Earth's surface, as a slight change in the acceleration constant $g$. Since the energy levels depend on $g$ as $E_n = (mg^2 \hbar^2)^{1/3} |\lambda_n|$, it follows that $\Delta E/E_n = {2\over 3}\Delta g/g$. Thus, a measurement performed, for instance, on the tenth energy level ($E_{10}=7.7 \,\rm peV$) with a given precision $\Delta E$ would yield a fivefold reduction of the relative error compared to the same measurement made on the ground state ($E_1=1.41 \,\rm peV$).

Finally, similar bouncing experiments have been proposed for antihydrogen atoms~\cite{Voronin_PRA11,Dufour_PRA13,Dufour2015} to measure the effect of gravity on antimatter. Our technique may also be useful to improve the accuracy of such experiments.

\begin{acknowledgments}
L. F. acknowledges the support of the Israel Science Foundation through grant 30/14.
T. J. and H. A. gratefully acknowledge support from the Austrian Science Fund~(FWF), Contract No.~P26781-N20, and from the French ANR, Contract No.~ANR-2011-ISO4-007-02 and FWF No.~I862-N20.
\end{acknowledgments}

\bibliography{qbouncer}

\end{document}